\documentclass{aastex}

\usepackage{graphicx}

\usepackage{emulateapj5}
\usepackage{times}

\usepackage{rotating}

\makeatletter

\newenvironment{inlinefigure}{%
\def\@captype{figure}%
\noindent\begin{minipage}{0.999\linewidth}\begin{center}}
{\end{center}\end{minipage}\smallskip}
\makeatother

\usepackage{mathptmx}

\def\Mo     {{\rm M}_{\odot}}

\begin{document}

%% LaTeX will automatically break titles if they run longer than
%% one line. However, you may use \\ to force a line break if
%% you desire.

\slugcomment{Accepted for publication in the Astrophysical Journal}

\title{Molecular Gas in the X-ray Bright Group NGC 5044 as Revealed by ALMA}

\author{Laurence P. David$^1$, Jeremy Lim$^2$, William Forman$^1$, Jan Vrtilek$^1$, Francoise Combes$^3$, Philippe Salome$^4$, Alastair Edge$^5$, Christine Jones$^1$, Ming Sun$^6$, Ewan O'Sullivan$^1$, Fabio Gastaldello$^7$, Pasquale Temi$^8$, Henrique Schmitt$^9$, Youichi Ohyama$^{10}$, Stephen Hamer$^5$, William Mathews$^{11}$, Fabrizio Brighenti$^{12}$, Simona Giacintucci$^{13}$, Sandro Bardelli$^{7}$ and Dinh-V Trung$^{14}$}
\affil{$^1$Harvard-Smithsonian Center for Astrophysics, 60 Garden St., Cambridge, MA 02138, USA}
\affil{$^2$Department of Physics, University of Hong Kong, Pokfulam Road, Hong Kong}
\affil{$^3$Observatoire de Paris, LERMA, CNRS, 61 Avenue de l’Observatoire, F-75014 Paris, France}
\affil{$^4$LERMA Observatoire de paris, CNRS, 61 rue de l’Observatoire, 75014 Paris, France}
\affil{$^5$Institute for Computational Cosmology, Department of Physics, Durham University, South Road, Durham DH1 3LE}
\affil{$^6$Department of Physics, University of Alabama in Huntsville, Huntsville, AL 35899, USA}
\affil{$^7$INAF - IASF-Milano, Via E. Bassini 15, 20133 Milano, Italy}
\affil{$^8$Astrophysics Branch, NASA/Ames Research Center, MS 245-6, Moffett Field, CA 94035}
\affil{$^9$Remote Sensing Division, Naval Research Laboratory, 4555 Overlook Avenue SW, Washington, DC 20375, USA}
\affil{$^{10}$Academia Sinica, Institute of Astromony and Astrophysics, Tawian}
\affil{$^{11}$University of California Observatoires/Lick Observatory, Department of Astronomy and Astrophysics, University of California, Santa Cruz, CA 95064, USA}
\affil{$^{12}$Dipartimento di Astronomia, Università di Bologna, via Ranzani 1, Bologna 40127, Italy}
\affil{$^{13}$Department of Astronomy, University of Maryland, College Park, MD 20742, USA}
\affil{$^{14}$Institute of Physics, Vietnamese Academy of Science and Technology, 10 DaoTan Street, BaDinh, Hanoi, Vietnam}

\begin{abstract}
An ALMA observation of the early-type galaxy NGC 5044, which resides
at the center of an X-ray bright group with a moderate cooling flow, detected 24
molecular structures within the central 2.5~kpc.  The masses of the molecular
structures vary from $3 \times 10^5$~M$_{\odot}$ to $10^7$~M$_{\odot}$ and
the CO(2-1) linewidths vary from 15 to 65~km~s$^{-1}$. Given the large
CO(2-1) linewidths, the observed structures are likely giant molecular associations
(GMAs) and not individual molecular clouds (GMCs). Only a few of the GMAs
are spatially resolved and the average density of these GMAs
yields a GMC volume filling factor of about 15\%.  The masses of the resolved
GMAs are insufficient for them to be gravitationally bound, however, the most massive
GMA does contain a less massive component with a linewidth of 5.5~km~s$^{-1}$
(typical of an individual virialized GMC). We also show that the GMAs cannot
be pressure confined by the hot gas.  Given the CO(2-1) linewidths of the
GMAs (i.e., the velocity dispersion of the embedded GMCs) they should disperse on a
timescale of about 12~Myr.  No disk-like molecular
structures are detected and all indications suggest that the molecular gas
follows ballistic trajectories after condensing out of the thermally
unstable hot gas.  The 230~GHz luminosity of the central continuum source
is 500 times greater than its low frequency radio luminosity
and probably reflects a recent accretion event.
The spectrum of the central continuum source also exhibits an absorption
feature with a linewidth typical of an individual GMC
and an infalling velocity of 250~km~s$^{-1}$.
\end{abstract}

%% Keywords should appear after the \end{abstract} command. The uncommented
%% example has been keyed in ApJ style. See the instructions to authors
%% for the journal to which you are submitting your paper to determine
%% what keyword punctuation is appropriate.

%% Authors who wish to have the most important objects in their paper
%% linked in the electronic edition to a data center may do so in the
%% subject header.  Objects should be in the appropriate "individual"
%% headers (e.g. quasars: individual, stars: individual, etc.) with the
%% additional provision that the total number of headers, including each
%% individual object, not exceed six.  The \objectname{} macro, and its
%% alias \object{}, is used to mark each object.  The macro takes the object
%% name as its primary argument.  This name will appear in the paper
%% and serve as the link's anchor in the electronic edition if the name
%% is recognized by the data centers.  The macro also takes an optional
%% argument in parentheses in cases where the data center identification
%% differs from what is to be printed in the paper.


\section{Introduction}

Cooling flow groups and clusters contain large amounts of hot
X-ray emitting gas that should be radiatively cooling on time
scales less than a Hubble time (see Fabian, Nulsen \& Canizares
1984 for an early review of cooling flows).  The primary uncertainty 
in the original cooling flow scenario was the ultimate fate of 
the cooling gas.  While diffuse H$\alpha$ emission was commonly found within the central 
dominant galaxy (CDG) in cooling flows (Heckman 1981; Hu et al. 1985)
and star formation was observed in cluster cooling flows as early as 
McNamara \& O'Connell (1989), the observed star formation rates 
were orders of magnitudes less than the inferred mass deposition rates
of the hot gas.  The cooling flow scenario underwent a major modification after 
the launch of the {\it Chandra} and {\it XMM-Newton} X-ray telescopes. 
{\it Chandra}, with its high spatial resolution, discovered 
AGN-induced cavities and shocks within cooling flows (e.g., McNamara et al. 2000; 
Blanton et al. 2003; Fabian et al. 2003; Forman et al. 2007; Randall et al. 2011), 
while both {\it XMM-Newton} and {\it Chandra} found little spectroscopic
evidence for large amounts of cooling gas (David et al. 2001; Peterson et al. 2003).  
By compiling observations of cooling flow clusters, several groups 
(Birzan et al. 2004; Dunn \& Fabian 2006, O'Sullivan et al. 2011) have shown
that the mechanical power of the AGN in the CDG is 
sufficient to prevent the bulk of the hot gas from cooling.  While AGN feedback 
probably prevents most of the hot gas from cooling, star formation 
has now been detected in many CDGs residing at 
the centers of cooling flows 
(e.g., Allen et al. 1995; Rafferty et al. 2006; Donahue et al. 2007; Quillen et al. 2008) 
and there are many indications that 
the star formation is a product of the cooling flow, including: 
1) a correlation between the star formation rate in CDGs and the spectroscopic 
mass deposition rate (O'Dea et al. 2008), 2) the existence of a sharp 
threshold for star formation that occurs when the cooling time of the 
hot gas is less than about 1~Gyr (Rafferty et al. 2008) or the entropy is less
than 30~keV~cm$^{2}$ (Voit et al. 2008) and 3) a correlation between the
star formation rate derived from {\it Spitzer} and {\it Herschel} data and 
the radiative cooling time of the hot gas (Egami et al. 2006a; Rawle et al. 2012).

Single dish CO surveys over the past decade have shown that a 
substantial fraction of CDGs in 
cluster cooling flows harbor molecular gas (Edge 2001; Salome et al. 2003).
Warm molecular gas has also been detected in CDGs by emission from
vibrationally excited molecular hydrogen (e.g., Jaffe \& Bremer 1997; 
Donahue et al. 2000; Egami et al. 2006b).
Recent Atacama Large (sub)Millimeter Array (ALMA) observations 
of A1664 (Russell et al. 2014) and
A1835 (McNamara et al. 2014) show that both of these systems 
contain more than ${10^{10} M_{\odot}}$ of molecular gas.
There is also evidence that A1835 may have a high velocity
molecular outflow driven by the radio jets or buoyant X-ray cavities.

In this paper we present the results of a cycle 0 ALMA observation
of NGC 5044 in the CO(2-1) emission line.  The NGC 5044 group of galaxies 
is the X-ray brightest group in the
sky and has a very smooth and nearly spherically symmetric large
scale X-ray morphology. However, the central region of NGC 5044
is highly perturbed with many AGN-inflated cavities, cool X-ray
filaments, cold fronts and multiphase gas (Buote et al. 2003; Gastaldello et al. 2009;
David et al. 2009; 2011).  NGC 5044 also hosts a system of very bright H$\alpha$
filaments (Caon et al. 2000; Sun et al. 2014).  
The presence of three cold fronts in the Chandra 
and XMM-Newton data (Gastaldello et al. 2013; O'Sullivan et al. 2014) and a 
peculiar velocity of 140~km~s$^{-1}$ relative to the group mean (Cellone \& Buzzoni 2005) 
suggest that NGC 5044 is likely undergoing a sloshing motion within the group center.
Large scale asymmetries in the X-ray morphology (Gastaldello et al. 2013) and 
regions of enhanced elemental abundances (O'Sullivan et al. 2014) suggest that 
the sloshing orbital plane of NGC 5044 is perpendicular to the plane of the sky based on 
the simulations in Roediger et al. (2011).

This paper is organized as follows. Section 2 contains a description
of the ALMA data reduction. Our results concerning the mass and 
kinematics of the observed molecular structures and correlations
between the molecular gas, dust, H$\alpha$ filaments and hot gas
are presented in $\S 3$.  Implications concerning the origin, 
dynamics and confinement of the molecular gas are discussed 
in $\S 4$ and our main results are summarized in $\S 5$. 

\bigskip

\section{Data Reduction}

We observed NGC 5044 with ALMA during Cycle 0 of the scientific observations.
A single pointing was made towards the center of NGC 5044
(RA=13:15:23.97, Dec-16.23.07.5) on 2012 January 13.  The primary beam 
of ALMA at 1.3~mm is $\sim$~27$^{\prime\prime}$ arcsec at full-width half maximum and 
provides a field of view of $\sim$~4.0 kpc at the distance of NGC 5044.    The 
observation spanned 1.0~hr, for a total on-source integration time of 29~min.
The quasar J1337-129, located $6.38^{\circ}$ from NGC 5044, served as the complex gain 
(secondary) calibrator. Scans of J1337-129 were made every 12 min bracketing scans 
of NGC 5044.  The quasar 3C 279 served as the bandpass calibrator and 
Titan as the absolute flux (primary) calibrator. 
The amount of precipitable water vapor during our observation was 1.4 mm

The correlator was configured to provide four spectral windows in two linear 
polarizations (no measurements of the cross products were made in ALMA cycle 0).  
Each spectral 
window spanned a total bandwidth of 1875~MHz and was split into 3840 channels 
so that each channel had a width of 488.28125~kHz.  One of the spectral windows 
was centered on the CO(2-1) line (1.3-mm band) at the systemic velocity 
of NGC 5044, providing a velocity resolution of 0.64~km~s$^{-1}$ in CO(2-1).
The other three spectral windows (229.336-231.211 GHz, 241.270-243.145 GHz
and 242.853-244.728 GHz) covered line-free regions to measure the 
continuum emission of NGC 5044, necessary for subtracting any continuum 
emission at the frequency of the CO(2-1) line.  As explained below, we detected 
relatively strong continuum emission from the central AGN of NGC 5044.

The data were calibrated using the software package CASA by the ALMA observatory.  
The calibrated data, along with a set of continuum-subtracted channel maps in CO(2-1), 
were delivered to us on 2013 March 4.  To make channel maps suitable to our own needs, 
we performed our own continuum subtraction in the CO(2-1) line and made channel 
maps at our own desired velocity resolutions and weightings of the visibility 
data.  The synthesized beam attained in our cycle 0 ALMA observation was 
1.4$^{\prime\prime}$ by 2.2$^{\prime\prime}$ FWHM (210 by 330~pc). 
We adopt a systemic velocity of 2758~km~s$^{-1}$ for NGC 5044 and a luminosity distance
of 31.2~Mpc (Tonry et al. 2001), which gives a physical scale 
in the rest frame of NGC 5044 of $1^{\prime\prime}=150$~pc.

\section{Results}

The most significant CO(2-1) feature in the ALMA Cycle 0 spectrum 
within the central 1~kpc diameter region is the
redshifted emission between 0 and 120~km~s$^{-1}$ with a sharp peak at 
60~km~s$^{-1}$ (see Fig. 1).  The blueshifted emission within this region is 
broader and less peaked, 
with emission spanning a velocity range from -300 to 0~km~s$^{-1}$. 
For comparison, we also show in Fig.1 the IRAM 30m CO(2-1) spectrum
and the ALMA CO(2-1) spectrum extracted from within a 11$^{\prime\prime}$
diameter region (i.e., the full-width half maximum of the IRAM 30m primary
beam). While the redshifted portions of the IRAM 30m and ALMA spectra are in reasonably
good agreement, the integrated flux density in the blueshifted portion of
the ALMA spectrum is only 20\% of that in the IRAM 30m data.
Similar results were found by Russell et al. (2014) 
for Abell 1664 and McNamara et al. (2014) for Abell 1835 when 
comparing IRAM 30m and ALMA data and is presumably due to the 
presence of diffuse CO(2-1) emission in these systems that
is resolved out in the ALMA data.

To search for discrete molecular structures in NGC 5044, we binned the ALMA data cube into 
velocity slices with bin widths ranging from 10 to 100~km~s$^{-1}$.  
We found that a bin width of 50~km~s$^{-1}$ optimized the signal-to-noise
of the emission lines.
%%Velocity slices smaller than 50~km~s$^{-1}$ reduce the signal from the 
%%molecular structures while larger velocity slices increase the noise.  
Fig. 2 displays channel maps between -350~km~s$^{-1}$ and 150~km~s$^{-1}$ in 
50~km~s$^{-1}$ slices.  A total of 24 molecular structures are detected with a 
CO(2-1) surface brightness exceeding $4 \sigma$ within the central 2.5~kpc 
(16.7$^{\prime\prime}$) where the primary beam response is greater than 0.3. More 
structures are detected at larger radii, but we limit all further analysis in 
this paper to these 24 molecular structures.
Our ALMA data covered a velocity range from -400 to 400~km~s$^{-1}$ centered

\begin{inlinefigure}
\center{\includegraphics[width=1.00\linewidth,bb=18 144 573 701,clip]{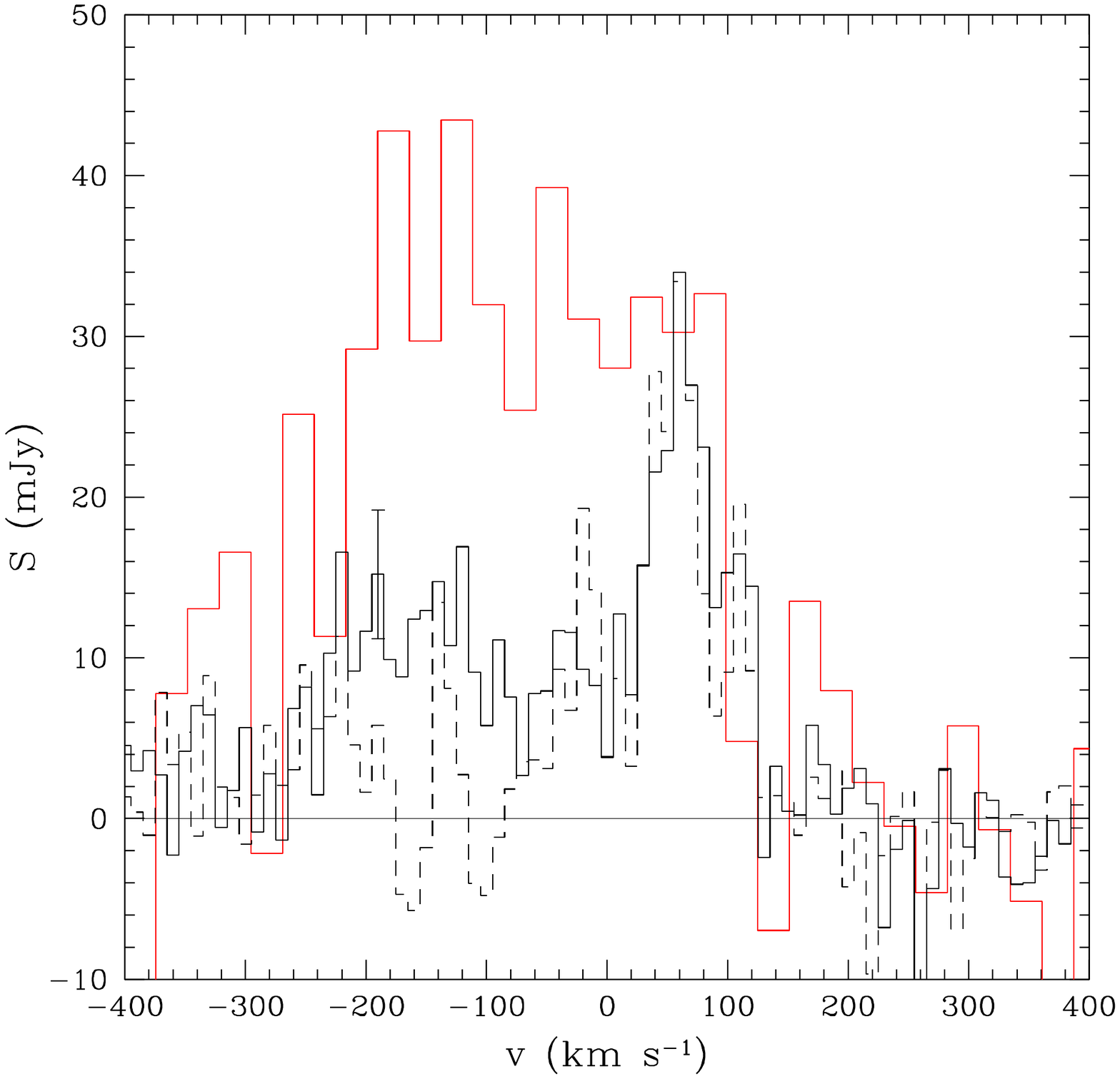}}
\caption{ALMA CO(2-1) spectrum from within the central
1~kpc ($6.7^{\prime\prime}$) diameter region (solid black line). For comparison,
the IRAM 30m CO(2-1) spectrum (solid red line) and the ALMA cycle 0 spectrum
(dashed black line) from within a 11$^{\prime\prime}$ diameter region
(i.e., the full-width half maximum of the IRAM 30m primary beam) are also shown.}
\end{inlinefigure}

\noindent
on the systemic velocity of NGC 5044.  No molecular structures were detected in 
emission with redshifts between 150-400~km~s$^{-1}$ or
blueshifts between 350-400~km~s$^{-1}$.  However, an absorption feature is detected 
in the central continuum source at 250~km~s$^{-1}$ and is discussed further in 
$\S 3.2$.  

The channel maps show that the velocity distribution of the 
molecular structures in NGC 5044 is highly asymmetric. There are 16 blueshifted 
structures and only 8 redshifted structures within the central 2.5~kpc. 
The blueshifted structures span a broad range in velocity, while the redshifted
structures are more localized in velocity space. The blueshifted structures
also are more centrally concentrated (at least in projection) compared to the 
redshifted structures.

CO(2-1) spectra were extracted for all 24 molecular structures identified in 
Fig. 2 and fit to a Gaussian profile.  The best-fit mean velocity, linewidth and integrated 
flux density are shown in Table 1 and the spectra are shown in Fig. 3.  For some 
molecular structures, the signal-to-noise was insufficient to obtain a well constrained fit.  
In addition, the spectra of some structures are not well described by a 
simple Gaussian.  For example, see the spectrum of molecular structure 13 in Fig. 3,
which has a extended blue wing.  For these molecular structures, only the mean velocity 
and integrated flux density are listed in Table 1.  The molecular mass was computed using 
the relation in Bolatto et al. (2013),

\begin{equation}
{M_{mol} = 1.05 \times 10^4~ S_{CO} \Delta v ~D_L^2 ~(1+z)^{-1} M_{\odot}}
\end{equation}

\noindent
where ${S_{CO} \Delta v}$ is the integrated CO (1-0) flux density in 
units of Jy~km~s$^{-1}$ and $D_L$ is the luminosity distance in units of Mpc.
This relation is based on the galactic CO-to-H$_2$ conversion factor of
$X_{CO}= 2 \times 10^{20}$~cm$^{-2}$~(K km s$^{-1}$)$^{-1}$.
We assume a CO(2-1) to CO (1-0) flux density ratio of 3.2 to estimate the 
molecular mass based on the observed CO(2-1) to CO (1-0) brightness 
temperature ratio of 0.8 for molecular clouds in spiral galaxies 
(e.g., Braine \& Combes 1992) and the factor of two in frequency.

The molecular masses of the structures range from $3 \times 10^5$~M$_{\odot}$ 
(corresponding to the $4 \sigma$ sensitivity limit) 
to $10^7$~M$_{\odot}$ and the linewidths vary from 
15 to 65~km~s$^{-1}$.  While resolved GMCs in the Milky Way and local 
group galaxies can have masses up to $10^7$~M$_{\odot}$, linewidths seldom 
exceed 10~km~s$^{-1}$ (Solomon et al. 1987; Blitz et al. 2006; 
Fukui et al. 2008; Bolatto et al. 2013) and we therefore refer to the 
molecular structures detected in NGC 5044 as giant molecular associations
(GMAs).

The total molecular mass of the GMAs listed in Table 1 is 
$5.1 \times 10^7$~M$_{\odot}$, the mean velocity is $-69.7 \pm 6.3$~km~s$^{-1}$ 
and the velocity dispersion is 122~km~s$^{-1}$, which is 
less than the stellar 
velocity dispersion of 237~km~s$^{-1}$.  While there is a significant difference 
in the number of red and blueshifted GMAs, the molecular mass is evenly divided
between the red and blueshifted GMAs.  This is due to the massive redshifted 
GMA 18 which contains 20\% of the total molecular mass.

\subsection{The central region}

There are several GMAs within the central 1~kpc diameter region that span 
more than one channel map. The channel maps suggest the possibility that GMAs 13
and 18 are contiguous across the central AGN, however the 
position-velocity diagram in Fig. 4 shows that GMAs 13 and
18 are distinct molecular structures.  Fig. 4 also shows that there
is no molecular structure with a smooth transition in velocities 
(from redshifted to blueshifted) across 
the systemic velocity of the galaxy, which would indicate 
the presence of a central disk.  To further investigate the 
kinematics of the central GMAs we generated velocity maps in two separate 
velocity slices (from -100 to 0~km~s$^{-1}$ and from 0 to 130~km~s$^{-1}$)
to prevent projection effects along the line of sight.  Fig. 5 shows that 
GMAs 11 and 13 have velocities near the 
systemic velocity of NGC 5044 at large 
radii and monotonically increasing blueshifted velocities close to the central AGN.
The blue shifted emission from GMA 13 is also visible in Fig. 3.
This suggests that these two GMAs are falling into the center of the galaxy
from the far side of the galaxy, similar to the situation observed in Perseus (Lim et al. 2008).
Fig. 6 shows that the gas velocity in GMA 18 increases along 
a SW to NE direction
and suggests that GMA 18 is falling into the central region of the galaxy from 
the near side of the galaxy. The feature in Fig. 6 toward the NE of GMA 18 at
a velocity of 20~km~s$^{-1}$ shows up as a separate molecular structure in Fig. 4.

The results shown in Table 1 were derived by fitting spectra
binned into 10~km~s$^{-1}$ channels.  GMA 18 produces 20\% of the total
flux within the central 2.5~kpc and the signal is strong enough to permit
a higher resolution study.  Fig. 7 shows a 1~km~s$^{-1}$ per channel spectrum
of GMA 18 which clearly shows the presence of two separate components.
Fitting a double Gaussian model yields a mean velocity, linewidth and molecular
mass for the narrow-line component of 58.7~km~s$^{-1}$, 5.5~km~s$^{-1}$ and
${M_{mol}=8.7 \times 10^5}$~M$_{\odot}$.  While the molecular mass of the narrow-line
component in GMA 18 is comparable with some of the lower mass GMAs detected in NGC 5044,
the linewidth is much smaller and more typical of an individual GMC (Solomon et al. 1987;
Blitz et al. 2006; Fukui et al. 2008; Bolatto et al. 2013).

Using the CASA task {\it imfit}, we fit 2D Gaussians to GMAs 11, 13 and 18
(the best resolved molecular structures in the center of NGC5044 which also
comprise 35\% of the total observed 

\newpage

\begin{inlinefigure}
\center{\includegraphics[width=2.1\linewidth]{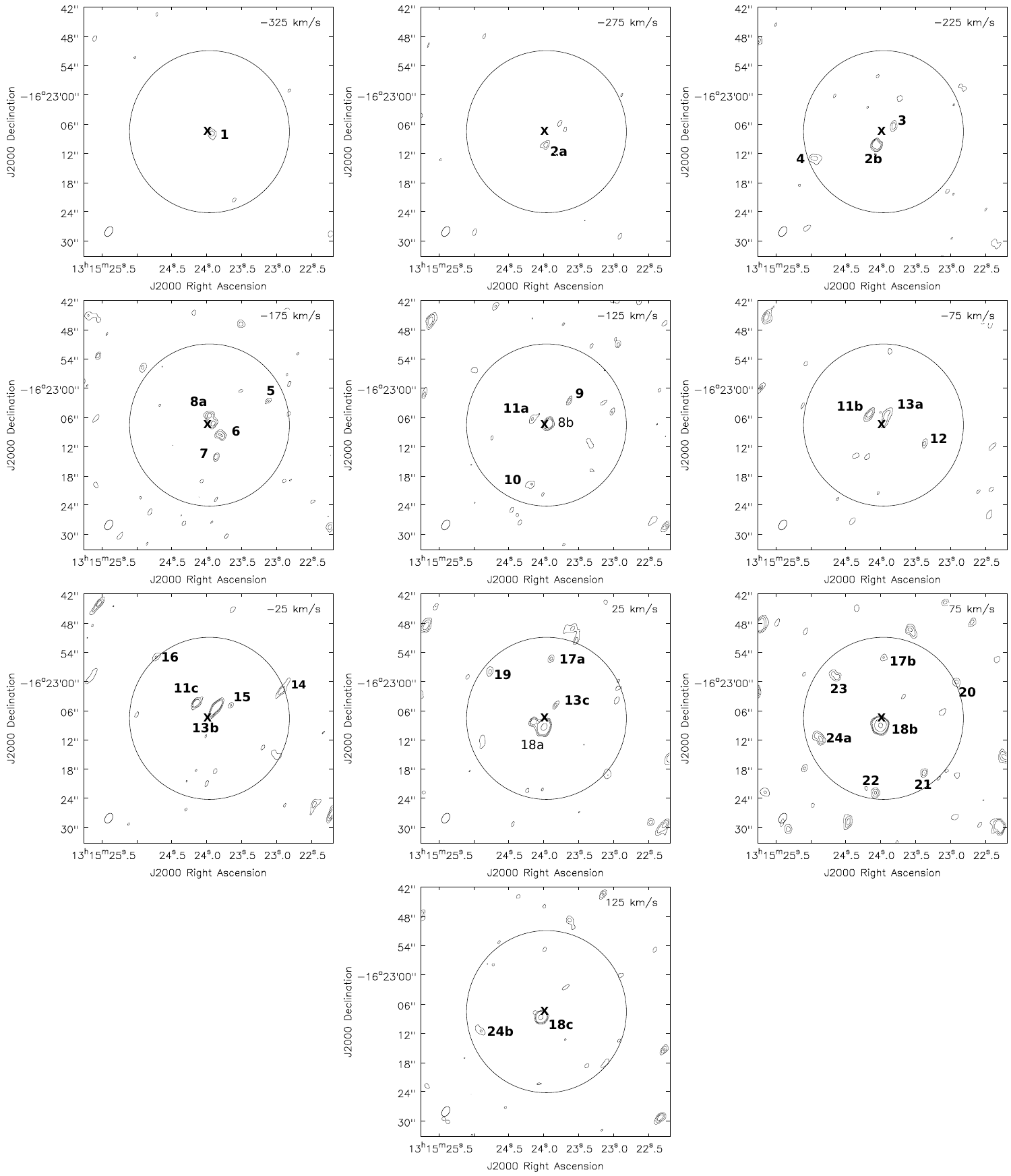}}
\caption{Contours of the integrated CO(2-1) intensity over a
velocity width of 50~km~s$^{-1}$ centered on the velocities given in the images.
Contours are shown at 3$\sigma$, 4$\sigma$, 5$\sigma$, $10 \sigma$ and $20 \sigma$,
where $\sigma=40$~mJy~beam$^{-1}$~km~s$^{-1}$. The circle corresponds to a
radius of 2.5~kpc in the rest frame of the galaxy.  The centroid of the continuum source
is marked with an "X".  The CO(2-1) emission from some GMAs is
contiguous across successive channel maps and these GMAs are labeled with
the same number and different letters on consecutive contour plots.}
\end{inlinefigure}

\newpage

\begin{inlinefigure}
\center{\includegraphics[width=2.0\linewidth]{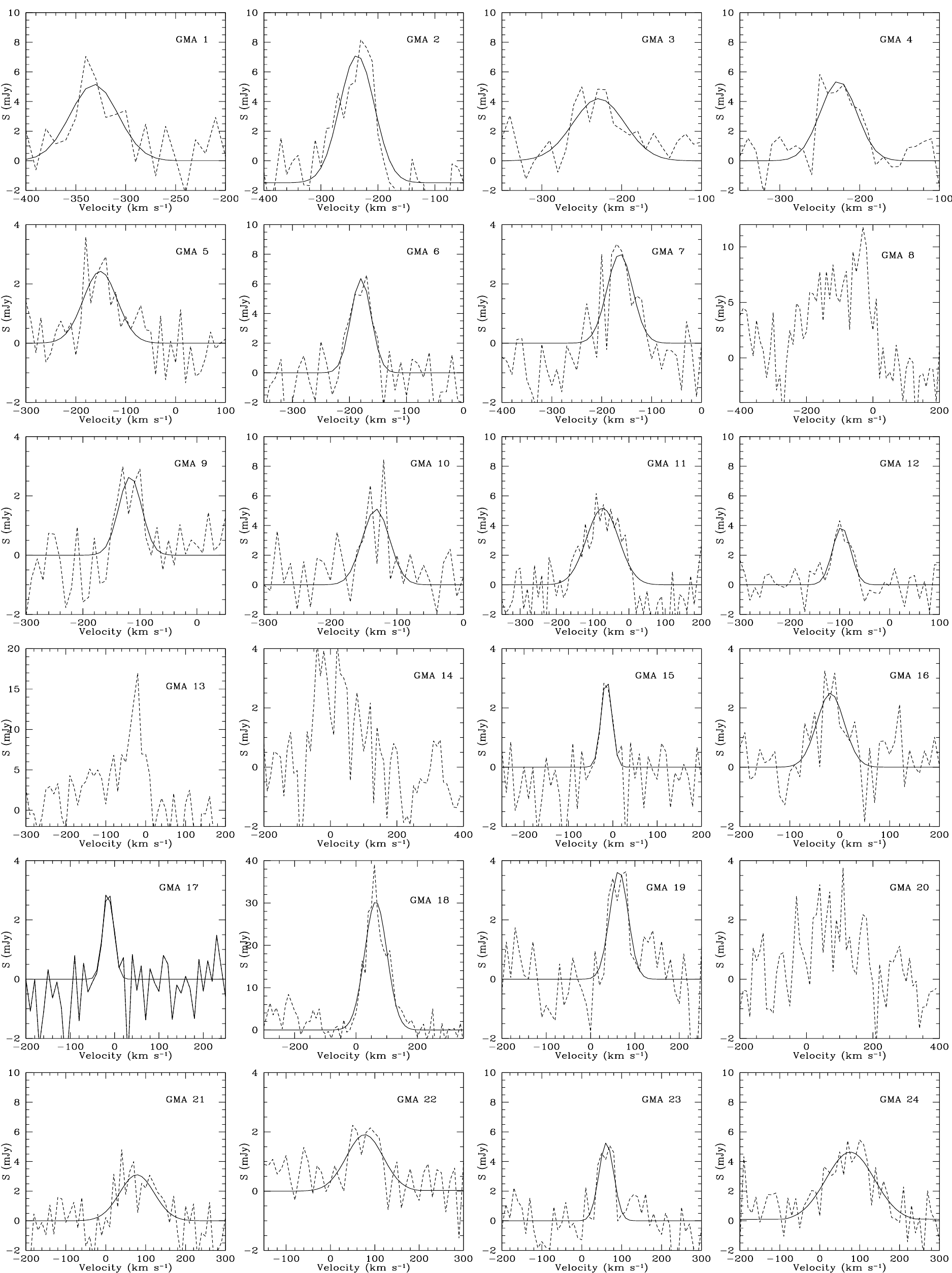}}
\caption{ALMA CO(2-1) spectra of the 24 GMAs along with the best-fit Gaussian profiles.}
\end{inlinefigure}

\newpage

\begin{inlinefigure}
\center{\includegraphics[width=1.00\linewidth,bb=30 119 417 759,clip]{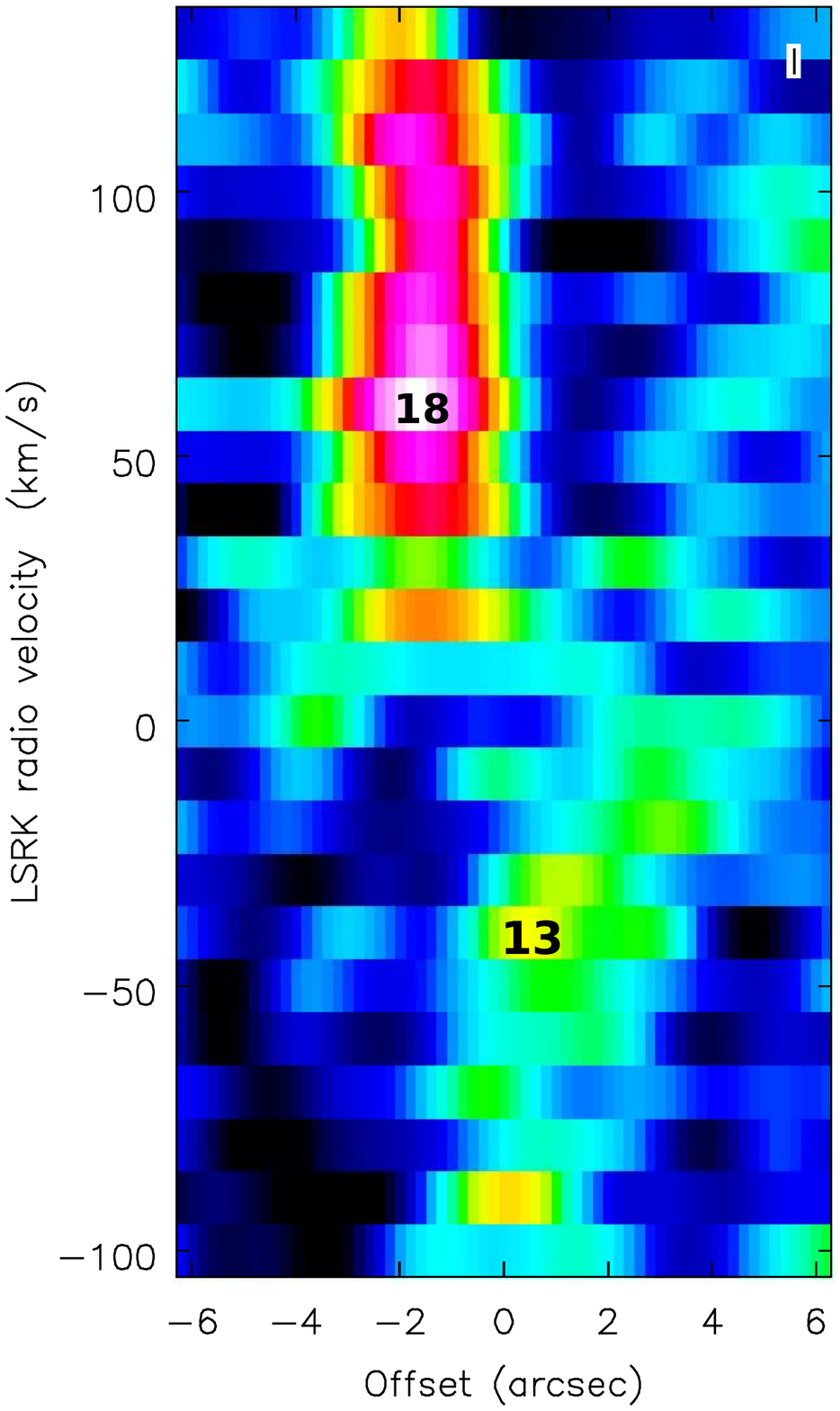}}
\caption{Position-Velocity diagram (where the horizontal axis corresponds to
the off-set in right ascension from the continuum source) for the central region of
NGC 5044. Note that GMAs 13 and 18 are distinct molecular structures. }
\end{inlinefigure}

\noindent
molecular mass). 
The task {\it imfit} computes the best-fit FWHM along the major and minor axes
of the 2D Gaussian.  We define the cloud radius as 
$r_c = \sqrt{\sigma_{maj} \sigma_{min}}$, where $\sigma=$~FWHM/2.35,
and obtain deconvolved cloud radii of $115 \pm 25$, $140 \pm 15$ and $120 \pm 12$~pc
for GMAs 11, 13 and 18, respectively. Using these radii and the molecular
cloud masses listed in Table 1, we obtain surface mass densities of
$100 \pm 30$, $53 \pm 10$ and $220 \pm 35$~ $M_{\odot}$~pc$^{-2}$, and 
${n_{H2}}$ volume densities of $13.0 \pm 4.7$, $5.8 \pm 1.2$ and 
$28.2 \pm 5.4$~cm$^{-3}$ for GMAs 11, 13 and 18, 
respectively. Assuming an
average density of $n_{H2}=100$~cm$^{-3}$ for true GMCs, the average GMC volume
filling factor for the three GMAs is 15\%.

The virial parameter, 

\begin{equation}
\alpha =5 \sigma^2 r_c / (G M_{mol})
\end{equation}

\noindent
can be used to determine if a cloud is self-gravitating ($\alpha=2$)
or self-gravitating and in virial equilibrium ($\alpha=1$).
For GMA 11 and the broad-line component of GMA 18, $\alpha=36$ and 24, respectively, 
indicating that these GMAs are not self-gravitating. We do not estimate $\alpha$
for GMA 13 since its spectrum does not follow a simple Gaussian profile.
While GMA 18 as a whole 

\begin{inlinefigure}
\center{\includegraphics*[width=1.00\linewidth,bb=26 306 436 599,clip]{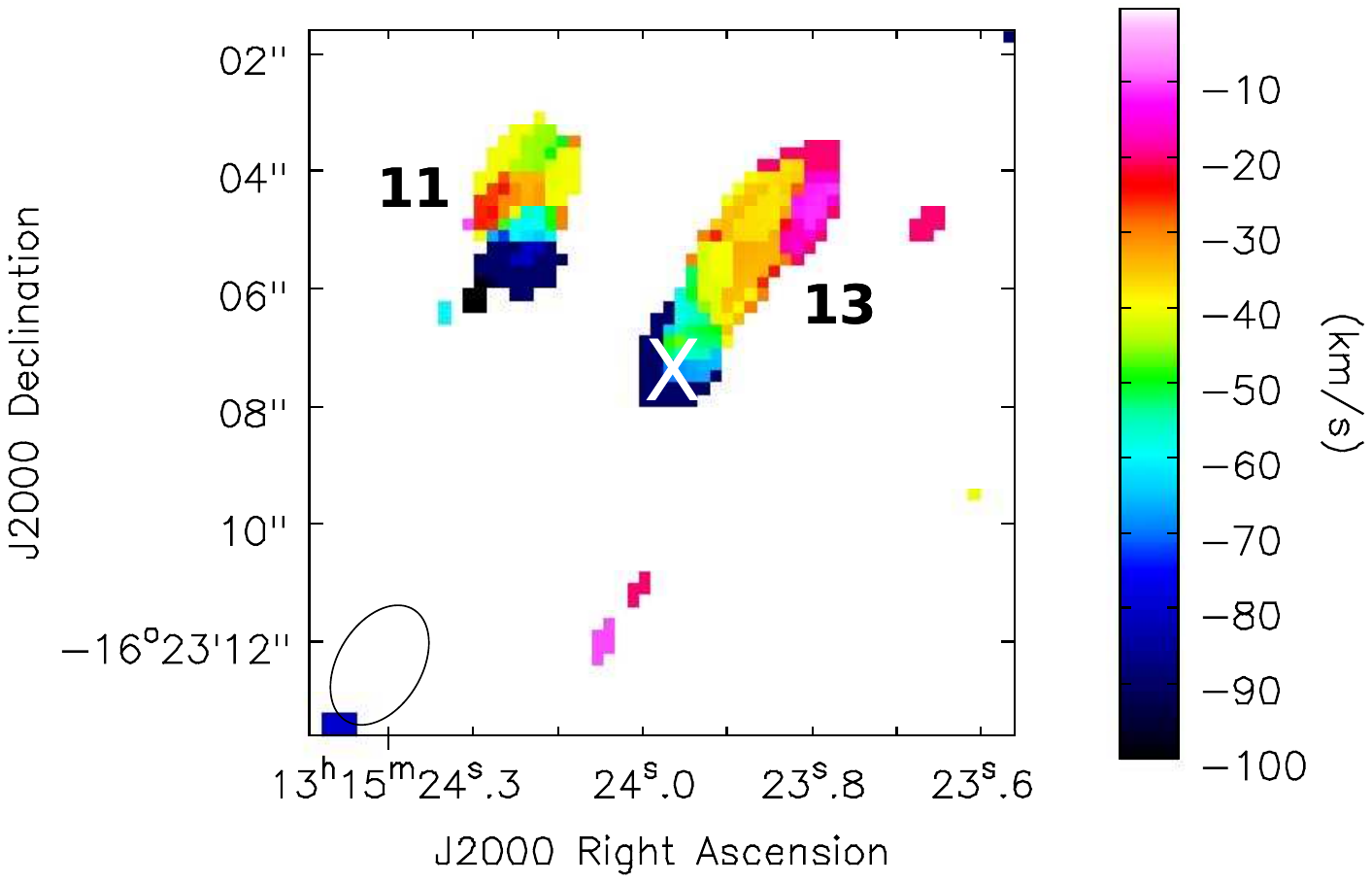}}
\caption{Intensity-weighted CO(2-1) mean velocity map between -100 and 0~km~s$^{-1}$.
The "X" marks the centroid of the continuum source. GMAs 11 and 13 (see Fig. 2) are
identified in the figure.}
\end{inlinefigure}

\noindent
is gravitationally unbound, the presence of the 
narrow-line feature seen in Fig. 7 shows that GMA 18 contains at least 
one virialized GMC.

\subsection{The Central AGN and Absorption Feature}

The central AGN in NGC 5044 is surprisingly bright in the ALMA data with a 
flux density of ${S_{\nu}= 55.3 \pm 3.9}$~mJy and luminosity of
${\nu L_{\nu} = 1.5 \times 10^{40}}$~erg~s$^{-1}$ measured in 
a line-free region.  We used the CASA task {\it imfit} to confirm that 
the central emission is consistent with a point source.  For comparison, the 
total flux densities (central point source plus extended emission) for 5044 at 
235~MHz and 610~MHz from GMRT data are 229~mJy and 39~mJy (Giacintucci et al. 2008). 
Thus, the 230 GHz luminosity of the central AGN is at least 
500 times greater than its luminosity at 610~MHz and a factor 
of a few greater than the 
bolometric X-ray luminosity (Giacintucci et al. 2008; David et al. 2009). 
As mentioned above, the large number of X-ray cavities within the central 
region of NGC 5044 are probably due to multiple AGN outbursts over the 
past $10^8$~yrs. The detection of 230~GHz continuum emission from the 
AGN shows that it is presently undergoing another outburst, probably
due to a recent accretion event.

It is unlikely that thermal emission from dust makes a substantial
contribution to the 230~GHz continuum emission. {\it Spitzer} data shows
that the 70~$\mu m$ dust emission is extended (Temi et al. 2007). 
In addition, even if all of the 70~$\mu m$ emission was assumed to 
originate from within the central region, the dust emission models
developed by Temi et al. (2007) predict a flux density 
at 230~GHz of less than 1~mJy.  The models in Temi et al. assume 
a steady-state balance between dust production and ion sputtering
and include heating from starlight and inelastic collision with 
thermal electrons. The FIR emissivity was computed by integrating over 
the local grain size (and therefore) temperature distribution.
%%Based on the analysis of { \it Spitzer} data, 
%%Amblard et al. (2014) give a dust temperature for NGC 5044 of 31.1~K

A pronounced absorption feature is seen in the CO(2-1) spectrum 
of the central continuum source (see Fig. 8.).  
Fitting a Gaussian profile to the absorption feature gives a mean velocity 
of $260.3 \pm 0.8$~km~s$^{-1}$ and a linewidth of $5.2 \pm 0.8$~km~s$^{-1}$.  
The line-of-sight velocity is a significant fraction of the circular velocity 
indicating that the cloud is falling into the central region of the galaxy on a 
nearly radial orbit.  The optical depth at line center is $\tau = 0.35$,
indicating that either 30\% of the continuum

\begin{inlinefigure}
\center{\includegraphics*[width=1.00\linewidth,bb=26 306 436 599,clip]{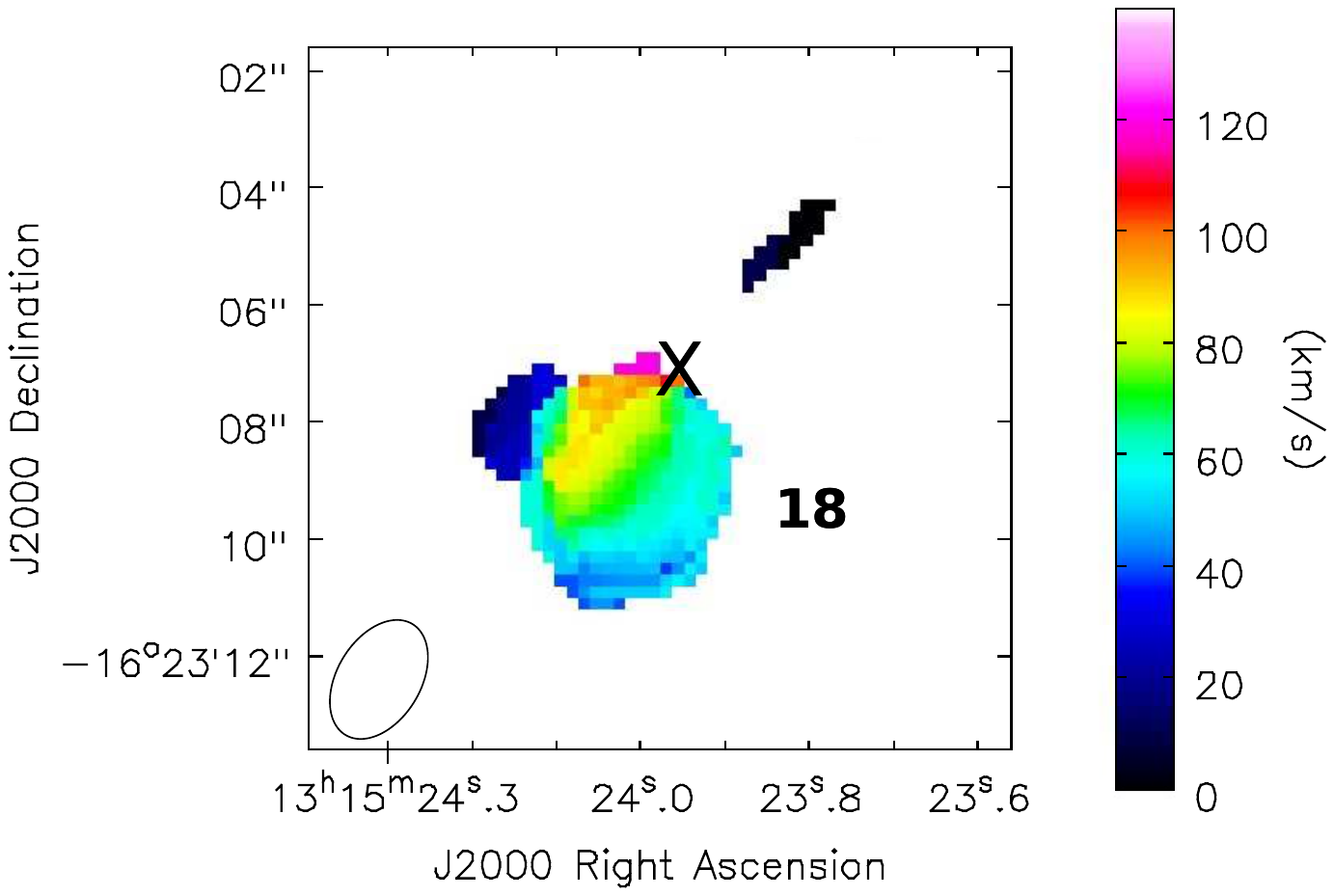}}
\caption{Intensity-weighted CO(2-1) mean velocity map between 0 and 150~km~s$^{-1}$.
The "X" marks the centroid of the continuum source. The most massive
molecular structure, GMA 18, is identified in the figure.}
\end{inlinefigure}

\noindent
source is covered by an
optically thick cloud or that the entire continuum source is covered by
diffuse molecular gas with $\tau = 0.35$.  
Due to the presence of faint line emission near the AGN, we were limited
to a maximum extraction region of 2$^{\prime\prime}$ by 2$^{\prime\prime}$ for 
the AGN spectrum shown in Fig. 8 which excludes some of the AGN continuum emission.
The use of a smaller aperture does not affect
the fraction of the continuum flux that is absorbed which is the main
parameter of interest.  We confirmed this by extracting spectra in several
smaller regions and found that 30\% 
of the continuum flux is 
absorbed at line-center in all spectra.
The linewidth of the absorber is typical of an individual GMC.
Assuming the linewidth-size relation in Solomon et al. (1987), we obtain 
a size for the absorbing 
cloud of 27~pc, implying that the radius of the continuum emission
is less than 50~pc. Alternatively, the absorption could be due 
to diffuse molecular gas that fully covers the continuum emission.
In this case, the column density of the diffuse CO must be 
${N_{CO}= 3.2 \times 10^{15}}$~cm$^{-2}$.  Since the ratio of ${N_{H2}}$ 
to ${N_{CO}}$ varies significantly in diffuse molecular gas 
(Burgh et al. 2007; Liszt 2012) we cannot accurately estimate ${N_{H2}}$.

\subsection{Correlations between the molecular gas, H$\alpha$ filaments,
dust and hot gas}

The distribution of the 24 GMAs listed in Table 1 is shown on an HST 
dust extinction map in Fig. 9.  This map was created based on
a F791W image obtained with the WFPC2 Planetary Camera.
The two-dimensional surface brightness distribution of the galaxy was
fit with two Sersic components using GALFIT (Peng et al. 2002; 2010).
The original image was divided by the best fitting model and 
resulting ratio image was converted to magnitudes.
While the most massive molecular structure, GMA 18 (the red circle close to the 
center of the figure), is spatially coincident with the highest extinction 
region toward the SE of the AGN, and a few of the GMAs trace the 
dust filaments toward the NW, there is not an obvious spatial correlation between 
the population of GMAs and the dust.  Most GMCs have visual extinctions significantly 
greater than 1~mag (Pineda et al. 2010), which would be easily identified
in Fig. 9. On theoretical grounds, it is difficult for CO to form with less shielding 
(Wolfire et al. 2010). It is possible that the GMAs not associated with regions of high 
extinction are located in dusty regions on the far side of the galaxy which would 
not produce much extinction. 

%%Temi et al. (2007) showed that the FIR emission from dust in 
%%NGC 5044 exceeds that expected from a steady-state balance between 
%%dust production in stellar winds and destruction by ion sputtering.  
%%The correlation between molecular gas and dust seen in Fig. 9 shows 
%%that at least some of the dust is shielded from the hot gas which 
%%will increase the lifetime and total mass of the dust.

\begin{inlinefigure}
\center{\includegraphics*[width=1.00\linewidth,bb=20 145 574 699,clip]{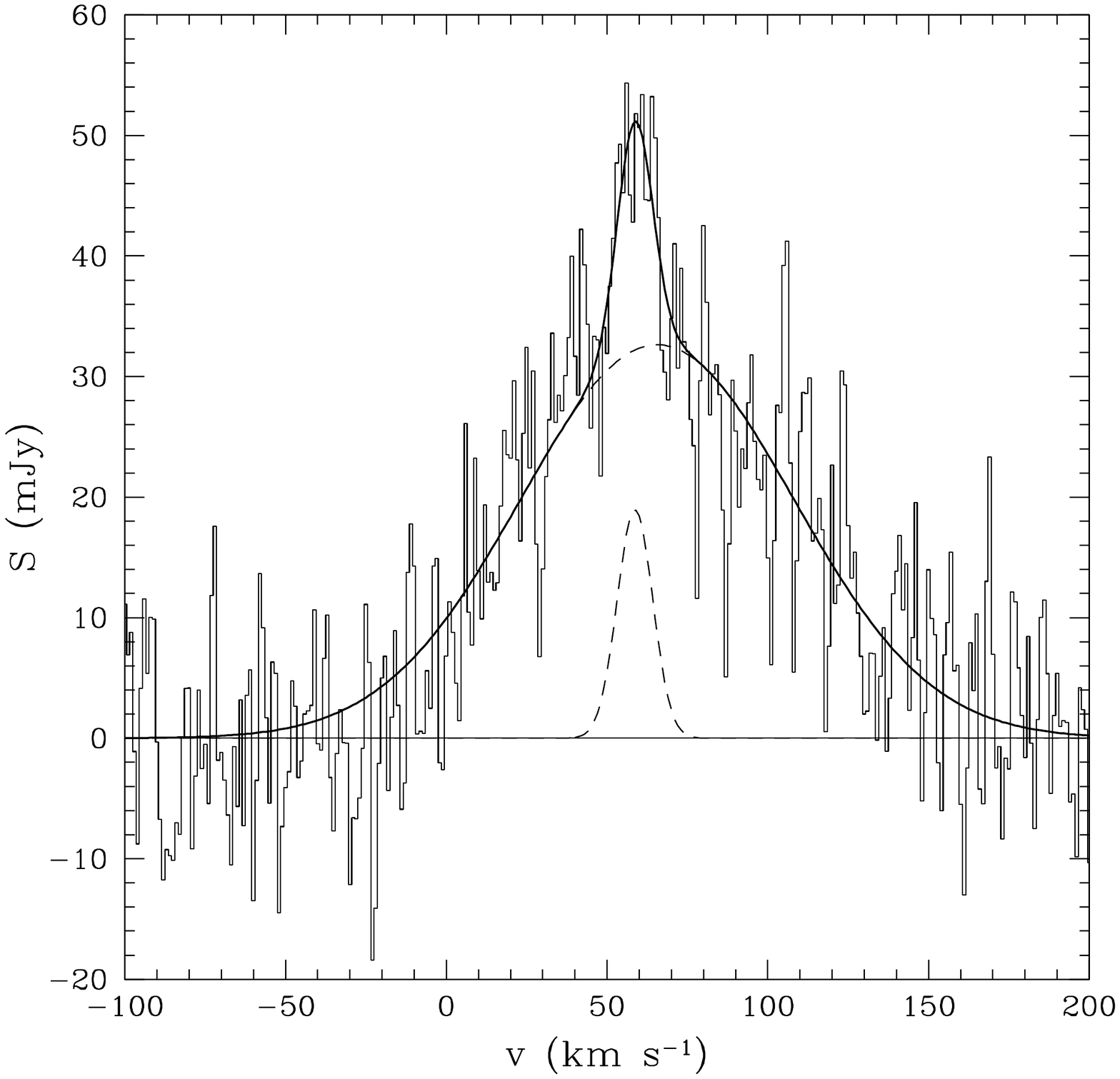}}
\caption{ALMA spectrum of GMA 18 along with the best-fit double Gaussian profile (solid
line).  The individual Gaussians are show as dashed lines.}
\end{inlinefigure}

NGC 5044 has the brightest system of H$\alpha$+[NII] filaments among 
groups of galaxies (Sun et al. 2014). The H$\alpha$+[NII] filaments
primarily extend along a north-south direction (see Fig. 10). Most of the 
GMAs are concentrated within the central region of 
the H$\alpha$+[NII] emission.  Due to the limited ALMA field-of-view,
it is difficult to determine how well the molecular gas traces 
the H$\alpha$+[NII] filaments at larger radii. 
Caon et al. (2000) measured the kinematics of the H$\alpha$+[NII]
filaments in NGC 5044 using long slit spectroscopy.  They found that the velocity 
profiles along the slits were fairly irregular, but the bulk of the 
H$\alpha$+[NII] emission was blueshifted by 60-100~km~s$^{-1}$ 
relative to the systemic velocity of NGC 5044. The predominance
of blueshifted H$\alpha$+[NII] emission is consistent with the 
presence of mostly blueshifted GMAs in the central region.

The unsharp masked 0.5-2.0~keV {\it Chandra} ACIS image from 
David et al. (2009) is shown in Fig. 11 along with the GMAs detected in the 
ALMA data.  Most of the GMAs 
are clustered within the central region which contains the coolest
and lowest entropy X-ray emitting gas.  The radiative cooling time
of the hot gas in the central region is about $4 \times 10^7$~yr, which 
is well below the observed threshold for triggering star formation
(Rafferty et al. 2008; Voit et al. 2008). The limited ALMA 
field-of-view does not permit a detailed comparison between the molecular gas 
and the larger scale structure seen in the ACIS image. 

Werner et al. (2014) recently presented {\it Herschel} observations of 
a sample of eight elliptical galaxies.  All six galaxies 
in their sample with H$\alpha$ emission were detected in [C II] (which is produced in the 
photodissociation region surrounding molecular clouds), including NGC 5044.  
The spatial resolution of {\it Herschel} at the [CII] emission line
is $12^{\prime\prime}$ and is insufficient to perform a detailed comparison
between the GMAs and the [CII] emission. Overall, the [CII] emission is 
elongated in a north-south direction, similar to the H$\alpha$ emission. 
The velocity structure of the [C II] emission is also
similar to the GMAs with predominately blueshifted emission 
relative to the systemic velocity of the galaxy near the center of 
NGC 5044 and mostly

\begin{inlinefigure}
\center{\includegraphics*[width=1.00\linewidth,bb=20 145 574 699,clip]{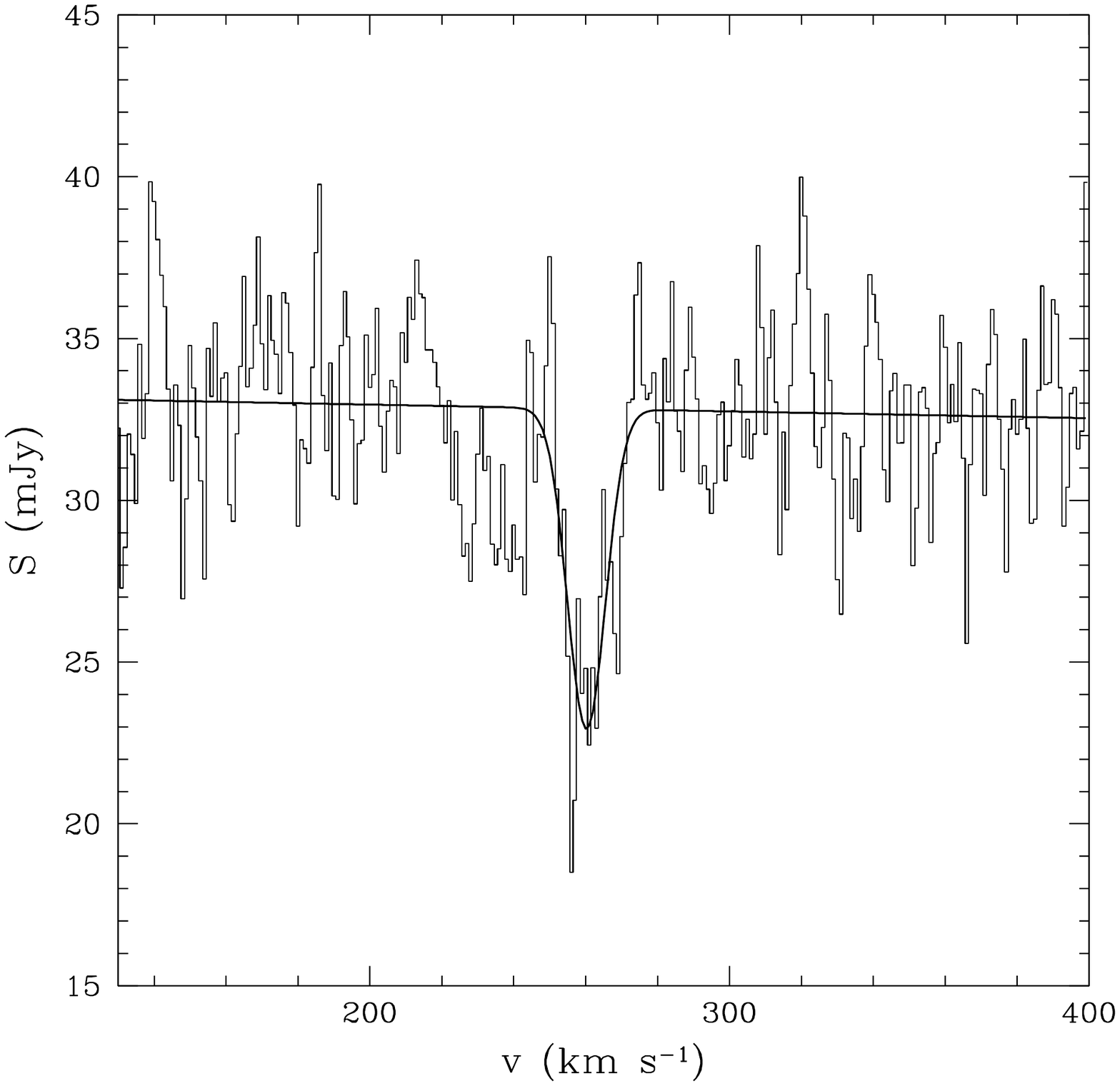}}
\caption{ALMA spectrum from within a $2^{\prime\prime}$ by $2^{\prime\prime}$
aperture centered on the AGN. Note the blueshifted absorption feature at
260~km~s$^{-1}$ with a linewidth of $5.2 \pm 0.8$~km~s$^{-1}$ which
is probably due to an infalling GMC.}
\end{inlinefigure}

\noindent
redshifted emission at larger radii.  However,
the [C II] emission is only detected over a velocity range of
about 120~km~s$^{-1}$ compared to a velocity range of 450~km~s$^{-1}$
for the CO(2-1) emission.

\section{Discussion}

\subsection{Source of the Molecular gas}

We detect 24 GMAs above a surface brightness limit of 
0.16~Jy~beam$^{-1}$ with a total mass of ${5.1 \times 10^7 M_{\odot}}$
within the central 2.5~kpc in NGC 5044. The molecular gas in 
NGC 5044 has a roughly azimuthally symmetric distribution, no evidence of any 
disk-like structures and a velocity dispersion less than the stellar 
velocity dispersion.  Werner et al. (2014) found that only systems like NGC 5044 
with thermally unstable gas, based on the Field criterion, contain [C II] emitting 
gas.  All these characteristics point to an intrinsic 
(i.e., cooling flow) and not extrinsic (i.e., merger with a gas rich system) 
origin for the molecular gas.  To estimate the mass shed from evolving stars we use 
the 2MASS data to compute ${L_K}$, a specific stellar mass loss rate of 
$\alpha=\dot M_*/M_*=5.4 \times 10^{-20}$~s$^{-1}$ 
(Renzini \& Buzzoni 1986, Mathews 1989) and a mass-to-light ratio 
of ${0.8 \Mo/L_{\odot K}}$ (based on dynamical measurements of 
early-type galaxies by Humphrey et al. 2006).  Within the central 10~kpc,
the stellar mass loss rate is ${\dot M_*}$= 0.23~M$_{\odot}$~yr$^{-1}$.  
There is evidence for multiphase hot gas out to at least 10~kpc in NGC 5044
(Buote et al. 2003; David et al. 2009), indicating that even in the
presence of multiple AGN outbursts, some hot gas is able to cool.  
The classical mass deposition rate 
(${\dot M_c = 2 \mu m_p L_x / 5 k T }$) within the central 10~kpc is 
5.1~M$_{\odot}$~yr$^{-1}$.  It is unclear what fraction of 
the X-ray emitting gas is able to cool due to AGN feedback. 
Fitting a cooling flow model to the ACIS data within the central region
of NGC5044 does not provide useful constraints on the spectroscopic 
mass deposition rate since the ambient gas temperature is only 0.7~keV
and the ACIS response declines rapidly below 0.5 keV.
The cavity power derived in David et al. (2009) is about one-half of the X-ray 
luminosity within the

\begin{inlinefigure}
\center{\includegraphics*[width=1.00\linewidth,bb=51 161 503 603,clip]{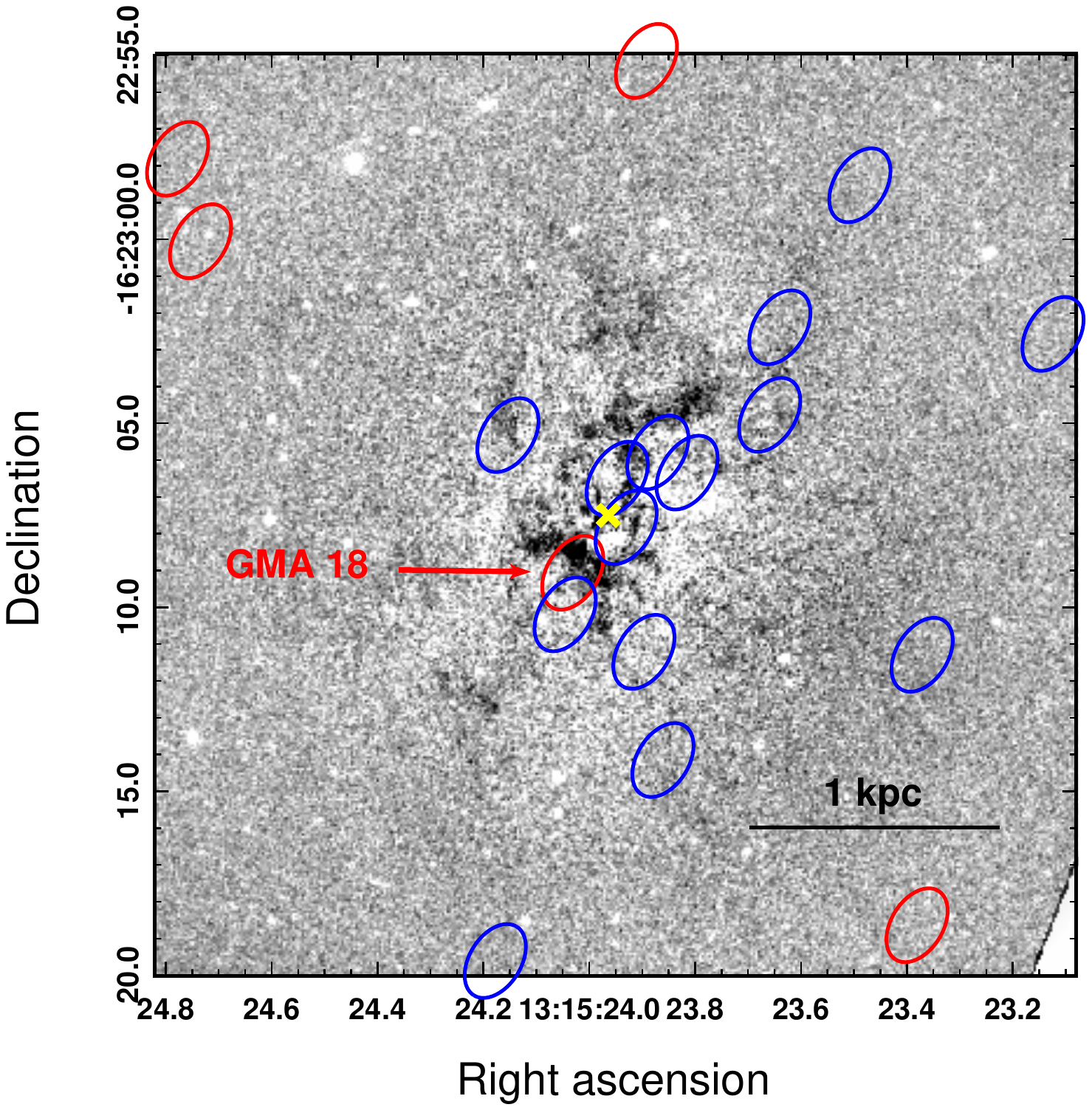}}
\caption{Location of the red and blueshifted GMAs on an HST dust extinction image.
The ellipses represent the ALMA beam in our observation.  The yellow "X" marks
the centroid of the continuum source.}
\end{inlinefigure}

\noindent
central 10~kpc, but this number depends on the 
uncertain volume of the cavities.  As long as 5\% of the hot gas is able to cool,
mass deposition from the cooling flow will be the dominant supply mechanism 
for the molecular gas.

Neglecting the suppression of gas cooling by AGN feedback, the 
molecular mass contained within the 24 GMAs can be produced
by gas cooling within the central 10~kpc in approximately 8~Myr.
Due to the limited sensitivity and coverage in our cycle 0 ALMA data, 
the 24 GMAs we detect is probably a lower limit. The H$\alpha$ and [C II] emission, 
which traces molecular gas, extend to at least 7~kpc.  The integrated CO(2-1) flux 
density in the IRAM 30m data is significantly greater than that in the ALMA data 
(especially in the blueshifted emission) which is presumably 
due to a significant component of diffuse molecular gas that is resolved out in the
ALMA data. Any additional molecular gas or 
reduction in the mass deposition rate due to AGN feedback would require 
a longer accumulation time for the molecular gas.

\subsection{Cloud Dynamics}

There are several indications of infalling molecular gas in
NGC 5044 (i.e., the absorption feature seen against the central 
continuum source and the velocity gradients in GMAs 11 and 13) and 
no indications of disk-like structures. 
The smooth X-ray morphology of the NGC 5044 group on
large scales indicates that the gas is in nearly hydrostatic
equilibrium, but the observed structure within the central region 
(see Fig. 11) suggests that there is some AGN-driven turbulence.
However, the turbulent velocities are probably a small fraction
of the sound speed of the hot gas.  Turbulent velocities greater than 
20-40~km~s$^{-1}$ would generate a heating rate due to the dissipation of 
turbulent kinetic energy that exceeds the radiative cooling rate 
of the hot gas (David et al. 2011).   The linewidth of an individual
GMA should reflect the turbulent velocity of the hot gas from which it
formed.  It is interesting to note that the observed linewidths of 
the GMAs in NGC 5044 are comparable to that required to balance radiative cooling
with the dissipation

\begin{inlinefigure}
\center{\includegraphics*[width=1.00\linewidth,bb=69 175 494 589,clip]{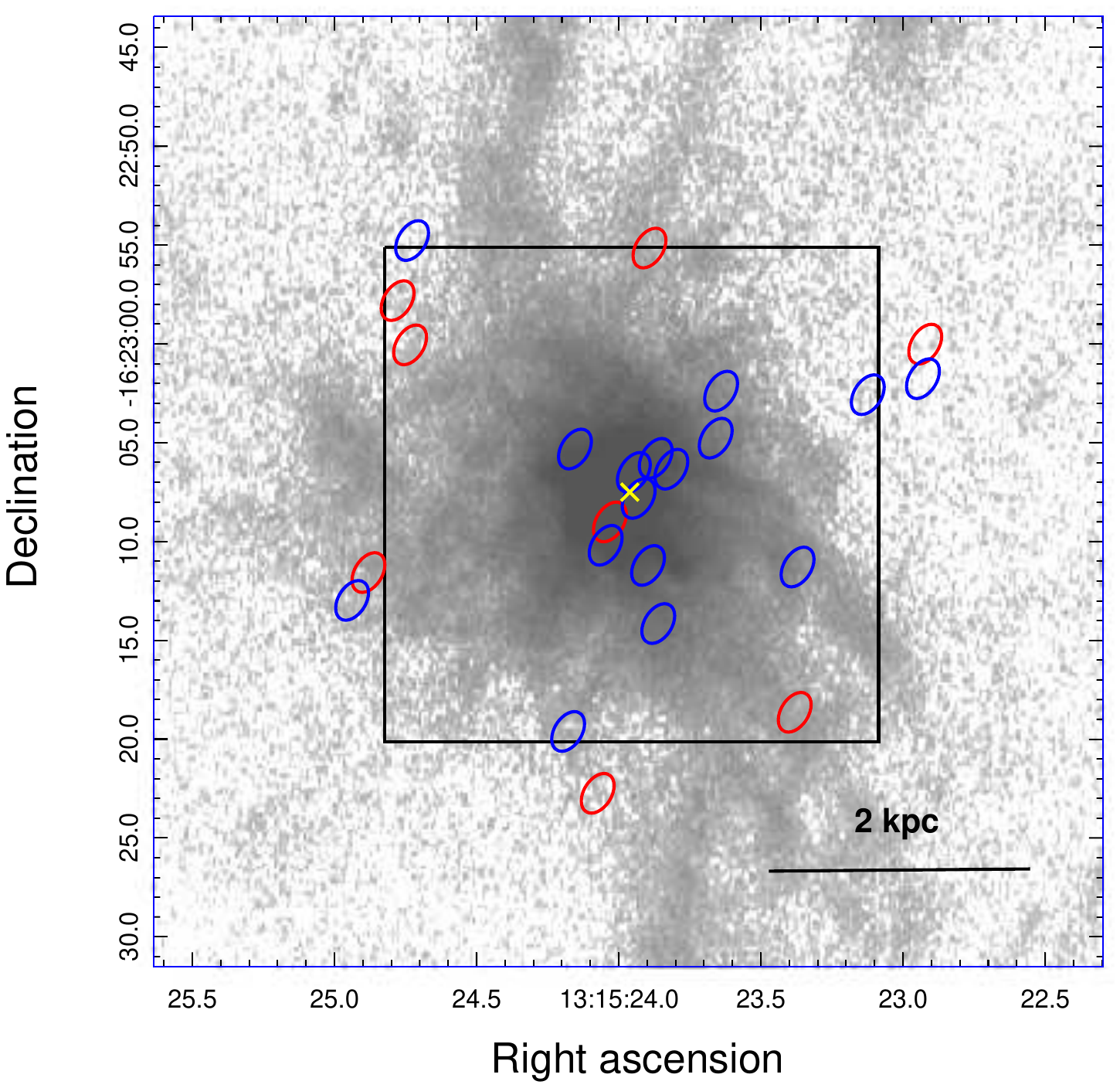}}
\caption{Location of the red and blueshifted GMAs on an H$\alpha$ image.
The black box shows the field-of-view of the HST image.}
\end{inlinefigure}

\noindent
of turbulent kinetic energy in the hot gas.

The kinematics of the molecular gas in NGC 5044 are very different from 
that observed in the more massive clusters A1835 
(McNamara et al. 2014) and A1664 (Russell et al. 2014).  ALMA 
cycle 0 observations show that both of these systems have more 
than ${10^{10} M_{\odot}}$ of molecular gas.  Abell 1664 has 
a massive central disk-like molecular structure and several high 
velocity clumps ($\sim 570$~km~s$^{-1}$) which could be either 
infalling or outflowing.  The molecular gas in A1835 is 
distributed between a nearly face-on disk and a bipolar outflow, 
possibly driven by radio jets or buoyant X-ray cavities (McNamara et al. 2014).

Both A1664 and A1835 host much more powerful AGNs compared to NGC 5044.  
The cavity power in NGC 5044 is ${P_{cav}=6 \times 10^{42}}$~erg~s$^{-1}$ 
(David et al. 2009), compared to ${P_{cav}=6-8 \times 10^{43}}$~erg~s$^{-1}$ 
in A1664 (Russell et al. 2014) and ${P_{cav}=10^{45}}$~erg~s$^{-1}$ in
A1835 (McNamara et al. 2014). 
%%Converting all of the cavity power in NGC 5044 
%%into turbulent kinetic energy within the molecular gas would produce a
%%an average CO(2-1) linewidth for the GMAs of 7~km~s$^{-1}$, which is significantly
%%less than the observed linewidths.
For buoyancy to have a significant impact on the molecular gas, the displaced 
mass in the X-ray cavities must equal the molecular gas mass.  For 
NGC 5044 this would require the displacement of all the hot gas 
within the central 2~kpc, which obviously has not happened (see Fig. 11). 
The total displaced gas mass in the cavities seen in Fig. 11 is approximately 
$8 \times 10^6 M_{\odot}$, which is only 15\% of the observed molecular gas mass.
While an X-ray cavity could still affect the dynamics of an 
individual GMA, there is no observed correlation between the velocities of
the GMAs and their proximity to X-ray cavities.  GMRT observations also show 
that NGC 5044 does not have a large scale collimated jet (Giacintucci et al. 2011).
Thus, due to the presence of only a weak AGN in NGC 5044, we conclude that 
the molecular gas essentially follows ballistic trajectories
after condensing out of the hot X-ray emitting gas.

\subsection{Are the GMAs confined?}

We showed above that GMAs 11 and 18 would have to be more than 
20 times more massive to be gravitationally bound.
While we cannot make a general statement about the entire GMA population in 
NGC 5044, it is clear that the largest GMAs

\begin{inlinefigure}
\center{\includegraphics*[width=1.00\linewidth,bb=54 158 497 586]{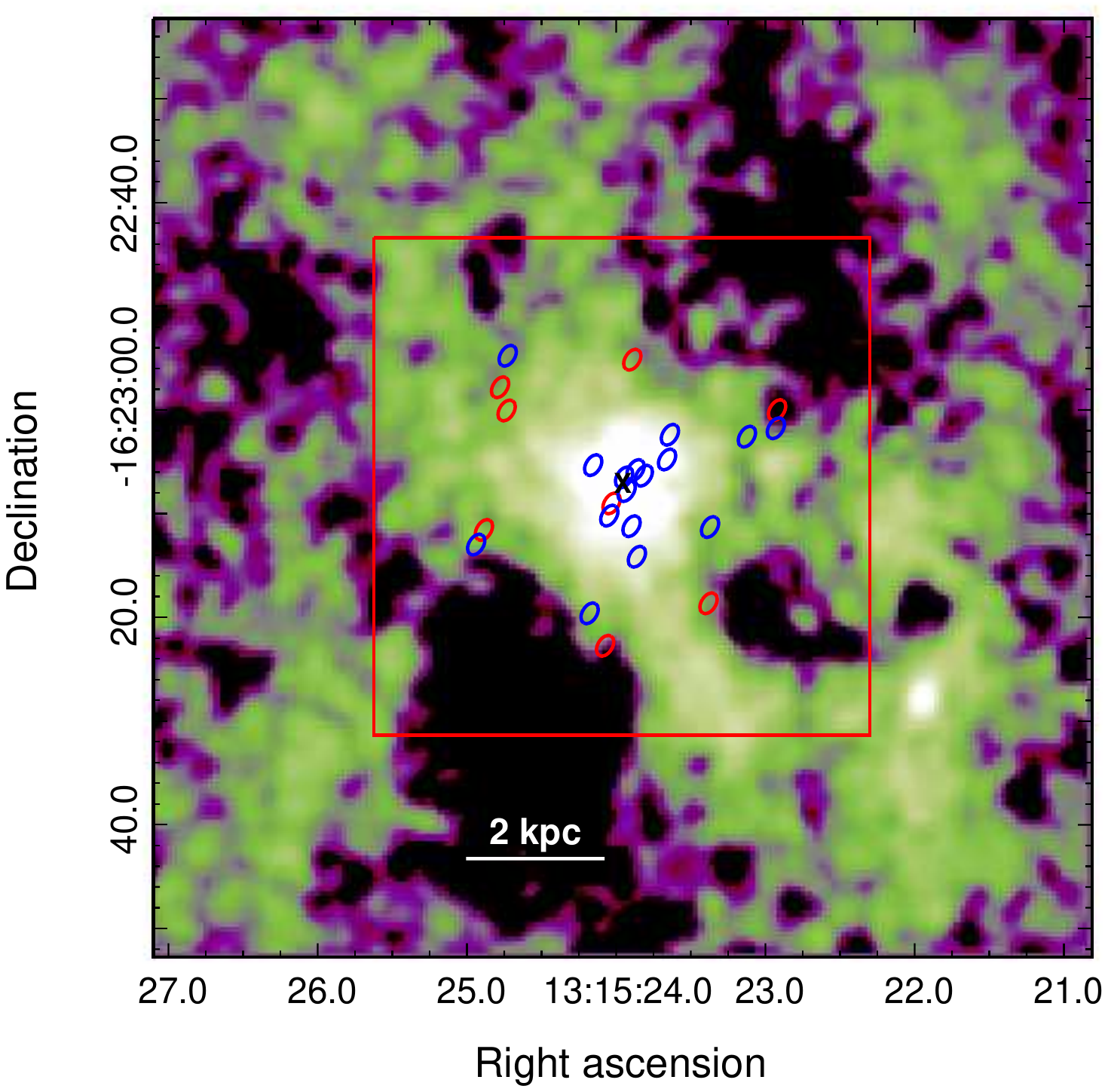}}
\caption{Location of the red and blueshifted GMAs on an unsharp masked ACIS image.
The red box shows the field-of-view of the H$\alpha$ image.}
\end{inlinefigure}

\noindent
are not gravitationally bound.
From the {\it Chandra} ACIS data we can compute the thermal pressure
of the X-ray emitting gas and determine if the GMAs are pressure confined. 
The thermal pressure of the hot gas in the vicinity of GMA 18 is 
${P_{ext}/k= 2.0 \times 10^6}$~K~cm$^{-3}$, assuming GMA 18 is located at 
a radial distance equal to its projected distance.
The CO(2-1) linewidth in GMA 18 is approximately 90 times the sound speed of 
the gas (assuming a kinetic temperature of 30~K), indicating that the pressure 
in GMA 18 is dominated by supersonic turbulence.  The turbulent pressure in 
GMA 18 is ${P_{turb}/k = \rho \sigma_v^2/k = 9.5 \times 10^6}$~K~cm$^{-3}$,
which is a factor of four greater than the thermal pressure of the
surrounding hot gas, suggesting that GMA is not pressure confined. 
However, even if the computed turbulent pressure was less than the thermal 
pressure of the hot gas, it is doubtful that GMA 18 could be confined
by the hot gas.
Our estimate for the turbulent pressure assumes that the bulk of GMA 18 is 
filled with gas at the volume-averaged density and that all of this gas has a 
turbulent velocity equal to the observed CO(2-1) linewidth.  Both of these 
assumptions are probably not valid. As mentioned above, the volume filling factor 
of the few resolved GMAs is fairly small, so the observed CO(2-1) linewidths should 
be thought of as the velocity dispersion of the embedded GMCs, each of which has 
a smaller internal velocity dispersion and higher density.  Due to the factor 
of $10^3$ contrast in densities between the embedded GMCs and the X-ray emitting 
gas, the thermal pressure of the hot gas will not decelerate and confine the 
ballistically moving GMCs within a GMA.  Thus, it appears unlikely that the 
few resolved GMAs in NGC 5044 are confined by either self-gravity or thermal pressure.

Individual GMCs will only be dynamically affected by the hot gas
after sweeping up a mass equal to their own mass.  Typical GMCs have
radii of a few pc and must therefore travel a few kpc due to the factor of $10^3$
contrast in density between the cold and hot gas before sweeping up their own mass. 
This distance is more than an order of magnitude greater than the 
largest observed GMA.  Since the GMAs are not confined by self-gravity or
thermal pressure, they should disperse in the time it takes a GMC to cross
the GMA. Given the observed 
radius and linewidth of GMA 18, embedded GMCs should traverse the GMA in 
a time of $t_{disp} \sim 2~r_c/\sigma$=12~Myr, which is less than the central 
40~Myr cooling time of the hot gas. This implies that the embedded GMCs within
a GMA must condense out of the hot gas at the same time and that the 
observed GMAs in NGC 5044 probably arise from local concentrations 
of thermally unstable parcels of hot gas.  One possible mechanism
for producing a local concentration of cooling parcels of hot
gas is the presence of dust, possibly transported outward
by AGN outbursts as discussed in Temi et al. (2007). 
Dust grains accelerate the cooling of the hot gas due to 
inelastic collisions between the thermal electrons and the dust
(Mathews and Brighenti et al. 2003). Reducing the cooling time of 
the gas also helps the non self-gravitating GMAs maintain their 
integrity as they cool.

\subsection{Effects of Star Formation}

The star formation rate in NGC 5044 based on {\it GALEX} and 
{\it WISE} data is 0.073~M$_{\odot}$~yr$^{-1}$ (Werner et al. 2014).
NGC 5044 also has unusual 
polycyclic aromatic hydrocarbon (PAH) line ratios compared
to other early-type galaxies (Vega et al. 2010) and a highly uncertain
stellar age (Marino et al. 2010), both of which may result from episodes 
of star formation over the past few Gyrs. The depletion time of the molecular 
gas due to star formation is approximately 700~Myr, which is typical 
for molecular gas in brightest cluster galaxies embedded in cooling flows 
(O'Dea et al. 2008; Voit \& Donahue 2011), but the IRAM 30m data shows
that there may be a significant amount of diffuse molecular gas
that was undetected in the ALMA data which would substantially increase 
the depletion time.

The Kennicutt (1998) relation between star formation rate and gas
surface density predicts a combined star formation rate from GMAs
11, 13 and 18 of 0.03~M$_{\odot}$~yr$^{-1}$. Considering that these
three GMAs comprise about 35\% of the observed molecular gas, this 
estimate is in reasonably good agreement with the observed star formation rate.
Assuming that one Type II supernova (SNeII) is produced for every 
100~M$_{\odot}$ consumed into stars and that each SNeII produces
$10^{51}$~erg, the SNeII heating rate in GMA 18 should be 
${H_{SN}= 6.7 \times 10^{39}}$~erg~s$^{-1}$.
If we assume that GMA 18 has a lifetime of 12~Myr (i.e., the crossing time 
of an individual 
GMC), then a total of $2.6 \times 10^{54}$~erg of energy will
be released by SNeII during this time, which corresponds
to 60\% of the turbulent kinetic energy in GMA 18.  Thus, depending
on the how SNeII energy is partitioned between radiative
and mechanical energy, SNeII could play a role in driving the 
internal turbulence and dissipation of the GMAs.

\section{Summary}

Our ALMA cycle 0 observation shows that the cooling flow
in the X-ray bright group NGC 5044 is a breeding ground for molecular gas.  
The CO(2-1) emission within the central 2.5~kpc is distributed among 24
molecular structures with a total molecular mass of $5.1 \times 10^7 M_{\odot}$.
Only a few of the molecular structures are spatially resolved and the 
observed radii, linewidths and masses of the these structures indicate that
they are not gravitationally bound. Given the large CO(2-1) linewidths of the
structures, they are likely giant molecular associations
and not individual clouds.  The average density of the few resolved
GMAs yields a GMC volume filling factor of 15\%.  
One narrow line component is detected within the largest GMA with 
a linewidth comparable to a virialized GMC.

While the large scale spatial distribution of the GMAs is azimuthally 
symmetric, there are some systematic differences in velocity space.  
Approximately 20\% of the molecular gas is contained in a single 
massive, redshifted GMA near the center of NGC 5044.  The remainder of
the molecular gas is divided between many lower mass GMAs with a
greater central concentration among the blueshifted GMAs compared
to the redshifted GMAs. On global scales, the kinematics of the molecular gas is 
similar to that observed in the H$\alpha$ filaments (Caon et al. 2000)
and [C II] emitting gas (Werner et al. 2014). 

There are no disk-like molecular structures observed in NGC 5044.  Two 
of the GMAs have velocity maps suggesting that they are falling into 
the center of NGC5044 from the far side of the galaxy.  There is also 
an absorption feature seen in the CO(2-1) spectrum of the central continuum 
source with an infalling velocity of 250~km~s$^{-1}$ and a linewidth typical of an 
individual GMC. Combining these results with the presence of infalling 
molecular gas in the Perseus cluster (Lim et al. 2008) suggest that 
infalling molecular gas is common in cooling flows.  The observed molecular 
mass and distribution and kinematics of the GMAs in NGC 5044 are consistent with 
a scenario in which the molecular gas condenses out of the thermally 
unstable hot gas and then follows ballistic trajectories.  
AGN-inflated buoyant cavities should not have a 
major impact on the dynamics of the molecular gas as a whole, except possibly 
in the immediate proximity of an X-ray cavity.

The integrated CO(2-1) flux density in the IRAM 30m observation of NGC 5044
is significantly greater than that produced by the molecular structures
detected in the ALMA data.  This is presumably due to a significant
component of diffuse molecular gas that has been resolved out in 
the ALMA data.  
%%The discrepency could also be due to the presence of a large 
%%number of widely separated molecular clouds (with typical radii of 10~pc)
%%whose CO(2-1) emission, once convolved with the ALMA beam (210 by 330~kpc),
%%falls below our $4 \sigma$ surface brightness detection threshold.  
%%Only clumps of GMCs will produce sufficient CO(2-1) emission, after being 
%%convolved with the ALMA beam, to exceed our detection threshold.
Due to the lack of gravitational or pressure confinement,
the GMAs detected in NGC 5044 should disperse on a timescale 
of about 12~Myr, which is less than the 40~Myr central cooling time of 
the hot gas, so the embedded GMCs within a GMA must condense out of 
the hot gas at the same time. Thus, the observed GMAs probably arise from local 
concentrations of over dense, thermally unstable regions in the hot gas.  

The wealth of information obtained from our 30 minute cycle 0 ALMA 
observation of the near by NGC 5044 group shows that it 
is an ideal target for investigating the formation of molecular 
gas in cooling flows and the long standing question about the ultimate 
fate of the cooling gas.  More sensitive and higher resolution ALMA observations 
in future cycles promise to illuminate the internal structure of the 
GMAs detected in the cycle 0 observation.

This paper makes use of the following ALMA data:
ADS/JAO.ALMA\#2011.0.00735.S. ALMA is a partnership of ESO (representing
its member states), NSF (USA) and NINS (Japan), together with NRC
(Canada) and NSC and ASIAA (Taiwan), in cooperation with the Republic of
Chile. The Joint ALMA Observatory is operated by ESO, AUI/NRAO and NAOJ.
The National Radio Astronomy Observatory is a facility of the National
Science Foundation operated under cooperative agreement by Associated
Universities, Inc.  This work was supported in part by NASA grant GO2-13146X.
ACE acknowledges support from STFC grant ST/I001573/1.
We would like to thank M. Birkinshaw for assistance in
learning CASA and analyzing the ALMA data and B. McNamara
and P. Nulsen for discussions about their cycle 0 ALMA observations
of clusters of galaxies.

\begin{table*}[t]
\begin{center}
\caption{Molecular Cloud Properties}
\begin{tabular}{lcccc}
\hline
ID & $<{v} >$ & $\sigma$ & ${S_{CO}\Delta v}$ & ${M_{mol}}$ \\
& (km~s$^{-1}$) & (km~s$^{-1}$) & (Jy~km~s$^{-1}$)  & ($10^5$~M$_{\odot}$)  \\
\hline
1 &  $-334.4 \pm 5.8  $&$ 21.0 \pm 7.4  $&$  0.186 \pm 0.057 $&$   5.94 \pm 1.82 $\\
2 &  $-235.4 \pm 5.5  $&$ 30.7 \pm 6.8  $&$  0.615 \pm 0.117 $&$  19.7 \pm 3.7 $\\
3 &  $-234.4 \pm 4.3  $&$ 24.1 \pm 4.9  $&$  0.236 \pm 0.043 $&$   7.53 \pm 1.37 $\\
4 &  $-226.4 \pm 3.6  $&$ 20.0 \pm 3.7  $&$  0.625 \pm 0.103 $&$   19.9 \pm 3.3 $\\
5 &  $-149.2 \pm 7.1  $&$ 36.3 \pm 8.4  $&$  0.417 \pm 0.084 $&$  13.3 \pm 2.7 $\\
6 &  $-177.6 \pm 3.2  $&$ 17.7 \pm 3.4  $&$  0.343 \pm 0.061 $&$  10.9 \pm 2.0 $\\
7 &  $-168.4 \pm 5.6  $&$ 37.5 \pm 6.5  $&$  0.342 \pm 0.052 $&$  10.8 \pm 1.7 $\\
8 &  $-146.1          $&$               $&$  1.19 \pm 0.14 $&$  38.1 \pm 4.7 $\\
9 &  $-112.7 \pm 7.6  $&$ 24.6 \pm 9.5  $&$  0.240 \pm 0.080 $&$   7.67 \pm 2.57 $\\
10 &  $-131.7 \pm 6.6  $&$ 19.1 \pm 7.1  $&$  0.286 \pm 0.097 $&$   9.14 \pm 3.10 $\\
11 &  $ -60.7 \pm 6.0  $&$ 53.0 \pm 7.5  $&$  1.31 \pm 0.15 $&$  41.8 \pm 4.7 $\\
12 &  $ -96.8 \pm 3.0  $&$ 20.3 \pm 3.4  $&$  0.371 \pm 0.055 $&$  11.8 \pm 1.8 $\\
13 &  $ -38.2          $&$                 $&$  1.03 \pm 0.12 $&$  32.9 \pm 3.9 $\\
14 &  $ -15.0          $&$                 $&$  0.300 \pm 0.097 $&$   9.58 \pm 3.09 $\\  
15 &  $ -13.8 \pm 3.8  $&$ 13.5 \pm 4.1  $&$  0.102 \pm 0.029 $&$   3.19 \pm 0.91 $\\
16 &  $ -21.5 \pm 5.8  $&$ 26.2 \pm 6.2  $&$  0.554 \pm 0.119 $&$  17.7 \pm 3.8 $\\
17 &  $  41.0 \pm 10.3 $&$ 45.9 \pm 13.1 $&$  0.546 \pm 0.135 $&$  17.4 \pm 4.3 $\\
18 &  $  58.9 \pm 2.7  $&$ 37.5 \pm 3.1  $&$  3.16 \pm 0.23 $&$  101.0 \pm 7.5 $\\
19 &  $  63.9 \pm 3.9  $&$ 20.7 \pm 3.9  $&$  0.302 \pm 0.103 $&$   9.65 \pm 3.30 $\\  
20 &  $  75.0          $&$               $&$  0.400 \pm 0.133 $&$  12.7 \pm 4.3 $\\  
21 &  $  84.5 \pm 12.2 $&$ 48.57 \pm 14.7 $&$  0.703 \pm 0.186 $&$  22.4 \pm 5.8 $\\
22 &  $  81.0 \pm 8.8  $&$ 29.7 \pm 12.3 $&$  0.284 \pm 0.101 $&$   9.10 \pm 3.22 $\\
23 &  $  61.2 \pm 3.3  $&$ 17.4 \pm 3.5  $&$  0.410 \pm 0.075 $&$  13.1 \pm 2.4 $\\
24 &  $  73.5 \pm 9.6  $&$ 63.9 \pm 13.1 $&$  2.11 \pm 0.37 $&$  67.5 \pm 12.0 $\\
\hline
\end{tabular}
\end{center}
\noindent
This table gives the GMA ID number, mean velocity, linewidth,
integrated CO(2-1) flux density and molecular mass. The S/N for 
GMAs 14 and 20 is insufficient to fit a Gaussian distribution. 
In addition, GMAs 8 and 13 do not have a simple Gaussian profile. 
No linewidths are listed for these GMAs.
\end{table*}


\begin{references}
\reference{}Allen, S. 1995, MNRAS, 276, 947.
\reference{}Amblard, A., Riguccini, L., Temi, P., Im, S., Fanelli, M. \& Serra, P. 2014, ApJ, 783, 135.
\reference{}Blanton, E., Sarazin, C. \& McNamara 2003, ApJ, 585, 227.
\reference{}Blitz, L., Fukui, Y., Kawamura, A., et al. 2007, in Protostars and Planets V, ed. B. Reipurth, D. Jewitt, \& K. Keil (Tucson, AZ: Univ. of Arizona Press), 81.
\reference{}Birzan, L., Rafferty, D., McNamara, B., Wise, M.\& Nulsen, P. 2004, ApJ, 607, 800.
\reference{}Braine, J. \& Combes, F. 1992, A\&A, 433, 443.
\reference{}Bolatto, A., Wolfire, M. \& Leroy, A. 2013, ARA\&A, 51, 207.
\reference{}Buote, D., Lewis, A., Brighenti, F. \& Mathews, W. 2003, 594, 741.
\reference{}Burgh, E., France, K. \& McCandliss, S. 2007, ApJ, 658, 454.
\reference{}Caon, N., Machetto, D. \& Pastoriza, M. 2000, ApJS, 127, 39.
\reference{}Cellone, S. \& Buzzoni, A. 2005, NNRAS, 356, 41.
\reference{}David, L., Nulsen, P. E. J., McNamara, B., Forman, W., Jones, C., Ponman, T., Robertson, B. \& Wise, M. 2001, ApJ, 557, 546.
\reference{}David, L., Jones, C., Forman, W., Nulsen, P., Vrtilek, J., O'Sullivan, E., Giacintucci, S. \& Raychaudhury, S. 2009, ApJ, 705, 624.
\reference{}David, L., O'Sullivan, E.,Jones, C., Giacintucci, S., Vrtilek, J., Raychaudhury, S., Nulsen, P., Forman, W., Sun, M. \& Donahue, M. 2011, ApJ, 728, 162.
\reference{}Donahue, M., Mack, J., Voit, G., Sparks, W., Elston, R. \& Maloney, P. 2000, ApJ, 545, 670.
\reference{}Donahue, M., Sun, M., O'Dea, C., Voit, G., Cavagnolo, K. 2007, AJ, 134, 14.
\reference{}Dunn, R. \& Fabian, A. 2006, MNRAS, 373, 959. 
\reference{}Edge, A. 2001, MNRAS, 328, 762.
\reference{}Egami, E., Misselt, K., Rieke, G., Wise, M., Neugebauer, G., Kneib, J., Le Floc'h, E., Smith, G., Blaylock, M., Dole, H., Frayer, D., Huang, J., Krause, O., Papovich, C., Pérez-González, P. \& Rigby, J. 2006a, ApJ, 647, 922.
\reference{}Egami, E., Rieke, G., Fadda, D. \& Hines, D. 2006b, ApJ, 652, 21.
\reference{}Fabian, A., Nulsen, P. E. J. \& Canizares, C. 1984, Nature, 310, 733.
\reference{}Fabian, A., Sanders, J, Allen, S., Crawford, C., Iwasawa, K., Johnstone, R., Schmidt, W. \& Taylor G. 2003, MNRAS, 344, 43.
\reference{}Forman, W. et. al. 2007, ApJ, 665, 1057.
\reference{}Fukui, Y. et al. 2008, ApJS, 178, 56. 
\reference{} Gastaldello, F., Buote, D., Temi, P., Brighenti, F., Mathews, W., Ettori, S. 2009, ApJ, 693, 43.
\reference{} Gastaldello, F. et al. 2013, ApJ, 770, 56.
\reference{}Giacintucci, S., Vrtilek, J., Murgia, M., Raychaudhury, S., O'Sullivan, E., Venturi, T., David, L., Mazzotta, P., Clarke, T., Athreya, R. 2008 ApJ, 682, 186.
\reference{}Heckman, T. 1981, ApJ, 250, 59.
\reference{}Hu, E., Cowie, L. \& Wang, Z. 1985, ApJSuppl, 59, 447.
\reference{} Humphrey, P., Buote, D., Gastaldello, F., Zappacosta, L., Bullock, J., Brighenti, F. \& Mathews, W. 2006, ApJ, 646, 899.
\reference{}Jaffe, W. \& Bremer, M. 1997, MNRAS, 284, 1.
\reference{}Kennicutt, R. 1998, ApJ, 498, 541.
\reference{}Larson, R. 1981, MNRAS, 194, 809.
\reference{}Lim, J., Ao, Y. \& Trung, D. 2008, ApJ, 672, 252.
\reference{}Liszt, H. \& Pety, J. 2012, A\&A, 541, A58.
\reference{}Marino, A., Rampazzo, R., Bianchi, L., Annibali, F., Bressan, A., Buson, L. M., Clemens, M., Panuzzo, P. \& Zeilinger, W. 2010, MNRAS, 411, 311.
\reference{}Mathews, W. 1989, AJ, 97, 42.
\reference{}Mathews, W. \& Brighenti, F. 2003, ApJ, 599, 992.
\reference{}Sun, M. et al. 2014 (in preparation).
\reference{}McNamara, B. et al. 2000, ApJ, 534, 135.
\reference{}McNamara, B. \& O'Connell, R. 1989, AJ, 98, 2018.
\reference{}McNamara, B. et al. 2014, ApJ, 785, 44.
\reference{}O'Dea, C. et al. 2008, ApJ, 681, 1035.
\reference{}O'Sullivan, E., Giacintucci, S., David, L., Gitti, M., Vrtilek, J., Raychaudhury, S. \& Ponman, T. 2011, ApJ, 735, 11.
\reference{}O'Sullivan, E., David, L. \& Vrtilek, J. 2014, MNRAS, 437, 730.
\reference{}Peng, C., Ho, L., Impey, C. \& Rix, H.-W 2002, AJ, 124, 266.
\reference{}Peng, C., Ho, L., Impey, C. \& Rix, H.-W 2010, AJ, 139, 2097. 
\reference{}Peterson, J., Kahn, S., Paerels, F., Kaastra, J., Tamura, T., Bleeker, J., Ferrigno, C. \& Jernigan, J. 2003, ApJ, 590, 207.
\reference{}Pineda, J., Goldsmith, P., Chapman, N., Snell, R., Li, D., Cambresy, L. \& Brunt, C. 2010, ApJ, 721, 686.
\reference{}Quillen, A. et al. 2008, ApJSuppl, 176, 39.
\reference{}Rafferty, D., McNamara, B. Nulsen, P. \& Wise, M. 2006, ApJ, 652, 216.
\reference{}Rafferty, D., McNamara, B. \& Nulsen, P. 2008, ApJ, 687, 899.
\reference{}Randall, S. Forman, W., Giacintucci, S., Nulsen, P., Sun, M., Jones, C., Churazov, E., David, L., Kraft, R., Donahue, M., Blanton, E., Simionescu, A. \& Werner, N. 2011, ApJ, 726, 86.
\reference{}Roediger E., Bruggen M., Simionescu A., Bohringer H., Churazov E., Forman W. R., 2011, MNRAS, 413, 2057
\reference{}Rawle, T. et al. 2012, ApJ., 747, 29.
\reference{}Russell, H. et al. 2014, ApJ, 784, 78.
\reference{}Salome, P. \& Combes, F. 2003, A\&A, 412, 657.
\reference{}Salome, P., Combes, F., Edge, A. C., Crawford, C., Erlund, M., Fabian, A. C., Hatch, N. A., Johnstone, R. M., Sanders, J. S. \&  Wilman, R. J. 2006, A\&A, 454, 437.
\reference{}Solomon, P., Rivolo, A., Barrett, J. \& Yahil, A. 1987, ApJ, 319, 730.
\reference{}Temi, P.,  Brighenti, F. \& Mathews, W. 2007, ApJ, 666, 222.
\reference{}Tonry, J., Dressler, A., Blakeslee, J., Ajhar, E., Fletcher, A., Luppino, G., Metzger, M. \& Moore, C. 2001, ApJ, 546, 681.
\reference{}Vega, O., Bressan, A., Panuzzo, P., Rampazzo, R., Clemens, M., Granato, G., Buson, L., Silva, L. \& Zeilinger, W. 2010, ApJ, 721, 1090.
\reference{}Voit, G. M., Cavagnolo, K., Donahue, M., Rafferty, D., McNamara, B., Nulsen, P. 2008, ApJ, 681, 5.
\reference{}Voit, G. M. \& Donahue, M. 2011, ApJ, 738, L24.
\reference{}Werner, N. et al. 2014, MNRAS, 439, 2291.
\reference{}Wolfire, M., Hollenbach, D. \& McKee, C. 2010, ApJ, 716, 1191.
\end{references}
\end{document}